\documentclass[12pt]{article}
\usepackage{graphicx}

\newcommand\pubnumber{SNSN-XXX-XX}
\newcommand\pubdate{\today}

\def\Title#1{\begin{center} {\Large #1 } \end{center}}
\def\Author#1{\begin{center}{ \sc #1} \end{center}}
\def\Address#1{\begin{center}{ \it #1} \end{center}}

\newcommand\pubblock{\rightline{\begin{tabular}{l} \pubnumber\\
         \pubdate  \end{tabular}}}
\newenvironment{Abstract}{\begin{quotation}  }{\end{quotation}}
\newenvironment{Presented}{\begin{quotation} \begin{center} 
             PRESENTED AT\end{center}\bigskip 
      \begin{center}\begin{large}}{\end{large}\end{center} \end{quotation}}

\usepackage{epsfig}
\usepackage{graphics}  
\usepackage{makeidx}
\usepackage{colordvi}
\usepackage{amsmath}
\usepackage{multirow}
\RequirePackage{xspace}
\usepackage{relsize}

\def\pep2{PEP-II \xspace}

\def\babar{\mbox{\slshape B\kern-0.1em{\smaller A}\kern-0.1em
    B\kern-0.1em{\smaller A\kern-0.2em R}}\xspace}

\def\invfb   {\ensuremath{\mbox{\,fb}^{-1}}\xspace}
\def\epem       {\ensuremath{e^+e^-}\xspace}

\def\CP                {\ensuremath{C\!P}\xspace}

\def\Y#1S{\ensuremath{\Upsilon{(#1S)}}\xspace}

\def\Bbar    {\kern 0.18em\overline{\kern -0.18em B}{}\xspace}

\def\BB      {\ensuremath{B\Bbar}\xspace} 

\def\Dz      {\ensuremath{D^0}\xspace}

\def\Kbar  {\kern 0.2em\overline{\kern -0.2em K}{}\xspace}

\def\Kz    {\ensuremath{K^0}\xspace}
\def\Kzb   {\ensuremath{\Kbar^0}\xspace}
\def\KzKzb {\ensuremath{\Kz \kern -0.2em - \kern -0.2em \Kzb}\xspace}
\def\KS    {\ensuremath{K^0_{\scriptscriptstyle S}}\xspace} 
\def\KL    {\ensuremath{K^0_{\scriptscriptstyle L}}\xspace} 
\def\KSKL {\ensuremath{\KS \kern -0.2em - \kern -0.2em \KL}\xspace}
\def\Kp    {\ensuremath{K^+}\xspace}
\def\Km    {\ensuremath{K^-}\xspace}
\def\Kpm   {\ensuremath{K^\pm}\xspace}

\def\pip  {\ensuremath{\pi^+}\xspace}
\def\pim  {\ensuremath{\pi^-}\xspace}
\def\pipm  {\ensuremath{\pi^\pm}\xspace}

\newcommand{\gevc}{\ensuremath{{\mathrm{\,Ge\kern -0.1em V\!/}c}}\xspace}
\newcommand{\mevc}{\ensuremath{{\mathrm{\,Me\kern -0.1em V\!/}c}}\xspace}
\newcommand{\gevcc}{\ensuremath{{\mathrm{\,Ge\kern -0.1em V\!/}c^2}}\xspace}
\newcommand{\mevcc}{\ensuremath{{\mathrm{\,Me\kern -0.1em V\!/}c^2}}\xspace}

\newcommand{\stat}{\ensuremath{\mathrm{(stat)}}\xspace}
\newcommand{\syst}{\ensuremath{\mathrm{(syst)}}\xspace}

\newcommand{\Dtokspi}{\ensuremath{D^{\pm}\to\KS\pipm}\xspace}
\newcommand{\Dtoksk}{\ensuremath{D^{\pm}\to\KS\Kpm}\xspace}
\newcommand{\Dstoksk}{\ensuremath{D_s^{\pm}\to\KS\Kpm}\xspace}
\newcommand{\Dstokspi}{\ensuremath{D_s^{\pm}\to\KS\pipm}\xspace}

\newcommand{\Dps}{\ensuremath{D_{(s)}}}
\newcommand{\Dpstoksk}{\ensuremath{\Dps^\pm \to \KS \Kpm}\xspace}

\begin{document}
\begin{titlepage}
\pubblock

\vfill
\Title{Search for \CP Violation in the Decays \Dtoksk, \Dstoksk, and \Dstokspi}
\vfill
\Author{Riccardo Cenci\\
(on behalf of the \babar collaboration)}
\Address{University of Maryland, College Park, Maryland 20742, USA}
\vfill
\begin{Abstract}
We report on a search for \CP violation in the decays \Dtoksk, \Dstoksk and \Dstokspi
using a data set corresponding to an integrated luminosity of $469\,\invfb$ 
collected with the \babar detector at the \pep2\ asymmetric energy \epem storage rings.
The \CP-violating decay rate asymmetries $A_{\CP}$ 
are determined to be
$(+0.13 \pm 0.36 \stat \pm 0.25 \syst)\%$, 
$(-0.05 \pm 0.23 \stat \pm 0.24 \syst)\%$, and 
$(+0.6 \pm 2.0 \stat \pm 0.3 \syst)\%$,
respectively for the three modes,
before correction for the relevant \KS asymmetry.
After this correction,
all the measurements are consistent with zero 
within one standard deviation.
These are currently the most precise measurements of these asymmetries.
\end{Abstract}
\vfill
\begin{Presented}
Charm 2012\\
5$^{\text{th}}$ International Workshop on Charm Physics\\
Honolulu, Hawai'i, USA, May 14$^{\text{th}}$--17$^{\text{th}}$, 2012
\end{Presented}
\vfill
\end{titlepage}
\def\thefootnote{\fnsymbol{footnote}}
\setcounter{footnote}{0}

\section{Introduction}

\CP violation (CPV) in charm decay is a sensitive probe of physics beyond the Standard Model (SM). 
Owing to its suppression within the SM, 
a significant observation of direct CPV in charm decay 
would indicate the possible presence of new physics effects in the decay process.   
In a previous publication~\cite{delAmoSanchez:2011zza}, 
we reported a precise measurement of CP asymmetry in the decay \Dtokspi, 
where the measured asymmetry was found to be consistent with the 
value expected for indirect CPV in the \Kz system. 
The LHCb and CDF collaborations have recently reported evidence 
for CPV in the decays $\Dz \to \Kp \Km$ and $\Dz \to \pip \pim$~\cite{Aaij:2011in,Collaboration:2012qw},
which, if confirmed, could be due either to new physics effects 
or to significant enhancement of penguin diagrams in charm decays~\cite{Isidori:2011qw,Franco:2012ck}.
Additional information and corroboration of these observations in other channels 
are necessary to resolve these questions.

In this report we present measurements of direct CPV asymmetries in the decay channels:
\Dtoksk, \Dstoksk, and \Dstokspi.
For these channels, as in the case of the decay \Dtokspi, 
we expect a $\Kz$-induced asymmetry of $\approx (\pm0.332\pm 0.006)\%$~\cite{Nakamura:2010zzi}. 
The sign of the  $\Kz$-induced
asymmetry is positive (negative) if a \Kz (\Kzb)
is present in the corresponding tree level Feynman diagram.
The exact magnitude of the 
asymmetry would depend on the requirements 
on the reconstructed $\KS\to\pi^+\pi^-$ decays and the decay kinematics~\cite{Grossman:2011zk}.  
Previous measurements of $A_{\CP}$ in these channels have been reported by 
the CLEO-c~\cite{:2007zt} 
and Belle collaborations~\cite{Ko:2010ng}.

For this analysis we employ a technique similar to that used 
in the measurement of CPV in \Dtokspi~\cite{delAmoSanchez:2011zza}, 
thus for the description of certain analysis details 
we refer to our previous publication.  

\section{Data Selection and Fit Procedure}

The data used for these measurements were recorded at or near the
$\Y4S$ resonance by the \babar detector at the \pep2 storage rings. 
The \babar detector and the coordinate system used throughout 
are described in detail in Refs.~\cite{Aubert:2001tu,Menges:2006xk}.
The total integrated luminosity used is $469\,\invfb$.
To avoid any potential biases in the measurements, we adopt a ``blind analysis'' approach, 
where for each channel we finalize the whole analysis procedure
prior to extracting $A_{\CP}$ from the data.

Signal candidates were reconstructed by combining a $\KS$ 
candidate reconstructed in the decay mode
$\KS\to\pi^+\pi^-$ with a charged pion or kaon candidate.
A \KS candidate is reconstructed from two oppositely charged
tracks with assigned the charged-pion mass.
\KS candidates are required to have an invariant mass within 
$\pm 10\,\mevcc$ of the nominal \KS mass~\cite{Nakamura:2010zzi},
which is equivalent to slightly more than $\pm 2.5\,\sigma$ in the measured
\KS mass resolution. The $\chi^2$ probability of 
the $\pi^+\pi^-$ vertex fit must be greater than $0.1\,\%$.
To reduce combinatorial background, we require the measured flight length
of the \KS candidate to be greater than 3 times its uncertainty.
A reconstructed charged-particle track that has $p_T\ge 400\,\mevc$ 
is selected as a pion or kaon candidate, where $p_T$ is the magnitude  
of the momentum in the plane perpendicular to the \epem collision axis.
At \babar, charged hadron identification is achieved through
measurements of specific ionization energy loss in the tracking system,
and the Cherenkov angle obtained from a detector of internally 
reflected Cherenkov light. A CsI(Tl) electromagnetic calorimeter 
provides photon detection, electron identification, and neutral pion 
reconstruction~\cite{Aubert:2001tu}. 
In our measurement, it is required that a pion candidate not be identified 
as a kaon, a proton, or an electron, while a kaon candidate is required
to be identified as a kaon and not as a pion, a proton, or an electron.
The criteria used to select pion or kaon candidates are very effective in reducing
the charge asymmetry from track reconstruction and identification, 
as inferred from the study of control samples, as described below.
A kinematic vertex fit to the whole decay chain is then performed,
and it is required that each candidate form a vertex 
close to the beam interaction region~\cite{Hulsbergen:2005pu}.
We retain only $\Dps^\pm$ candidates having a $\chi^2$ probability for this fit greater than 0.1\%,
and an invariant mass $m(\KS h), h = \pi, K, $ within $\pm65\mevcc$ 
of the nominal $\Dps^+$ mass~\cite{Nakamura:2010zzi};
this is equivalent to more than $\pm 8\,\sigma$ in the measured
$\Dps^\pm$ mass resolution.

We further require the magnitude of the $D^\pm_{s}$ candidate momentum in the 
$\epem$ center-of-mass (CM) system, $p^*(D^\pm_{s})$, 
to be between 2.6 and 5.0 \gevc to suppress the combinatorial background from \BB events.
For the decay channel \Dtoksk, which suffers from lower statistics, 
we retain candidates with $p^*(D^\pm)$ between 2.0 and 5.0 \gevc.
This selects some signal events originating from $B$ meson decays~\cite{CPTinv},
while maintaining an acceptable level of combinatorial background. 
Additional background rejection is obtained by requiring that the impact parameter 
of the $\Dps^\pm$ candidate with respect to the \epem interaction region~\cite{Aubert:2001tu}, 
projected onto the plane perpendicular to the collision axis,
be less than 0.3 cm, and that the $\Dps^\pm$ lifetime $\tau_{xy}(\Dps^\pm)$ be between $-15$ and $35$ ps. 
The lifetime is measured using $L_{xy}(\Dps^\pm)$, defined as the distance of 
the $D^\pm_{(s)}$ decay vertex from the interaction region
projected onto the plane perpendicular to the collision axis.

In order to further optimize the sensitivity of the $A_{\CP}$ measurements, 
a multivariate algorithm is constructed from seven 
discriminating variables for each $\Dps^\pm$ candidate.
These are: $\tau_{xy}(\Dps^\pm)$, $L_{xy}(\Dps^\pm)$,
$p^*(\Dps^\pm)$, the momentum magnitude and transverse component 
with respect to the beam axis for the \KS, and for the pion or kaon candidate.
For \Dtoksk and \Dstoksk the multivariate algorithm with the best performance is a Boosted Decision Tree (BDT),
while for \Dstokspi the best algorithm is a Projective Likelihood (LH) method~\cite{Speckmayer:2010zz}.
The final selection criteria, based on the output of the relevant multivariate selector,
are optimized using the $S/\sqrt{S+B}$ ratio as defined in Ref.~\cite{delAmoSanchez:2011zza}.

\begin{figure}[tb]
\begin{center}
\begin{tabular}{ccc}
\includegraphics[width=0.31\textwidth]{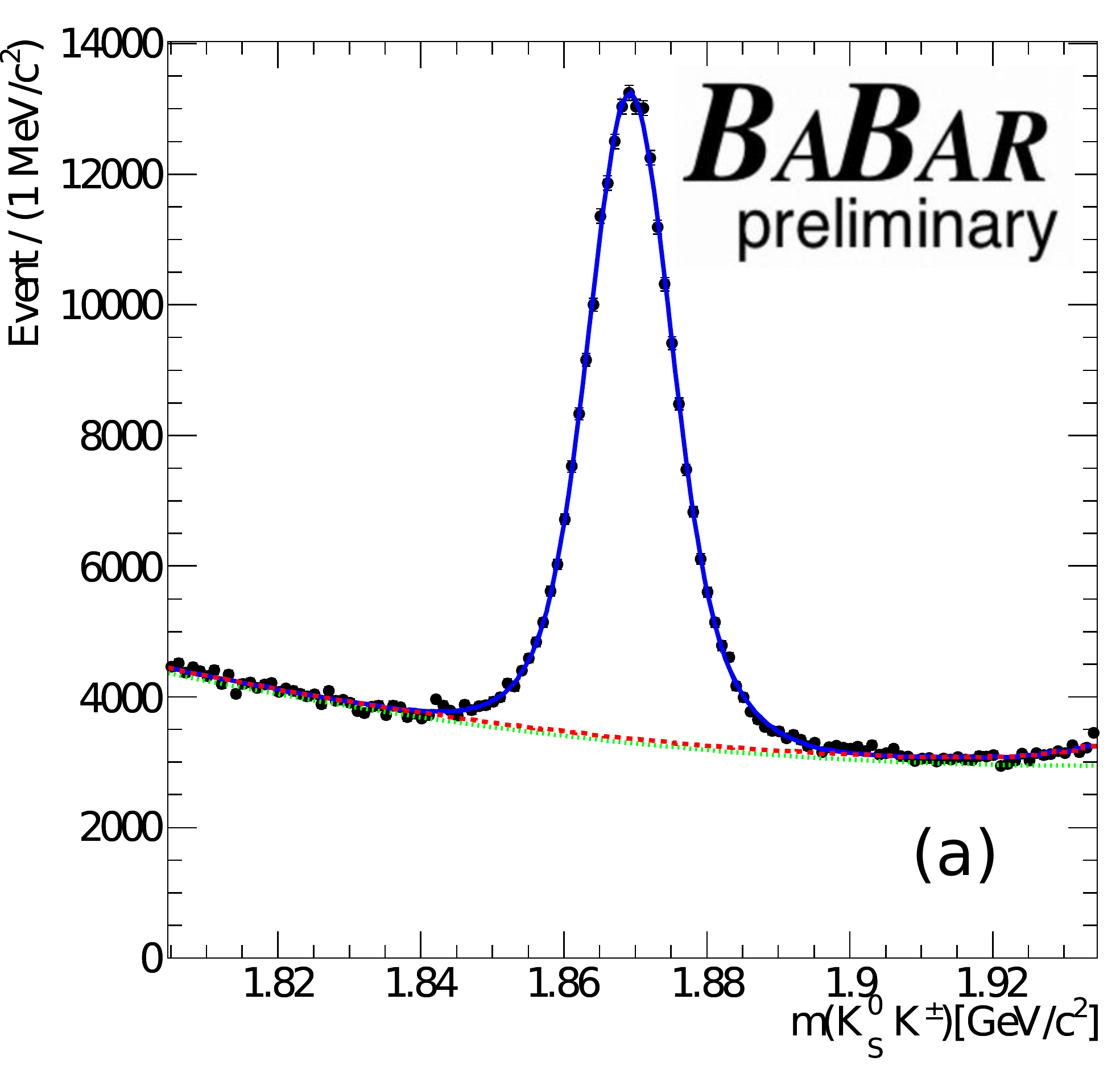} &
\includegraphics[width=0.31\textwidth]{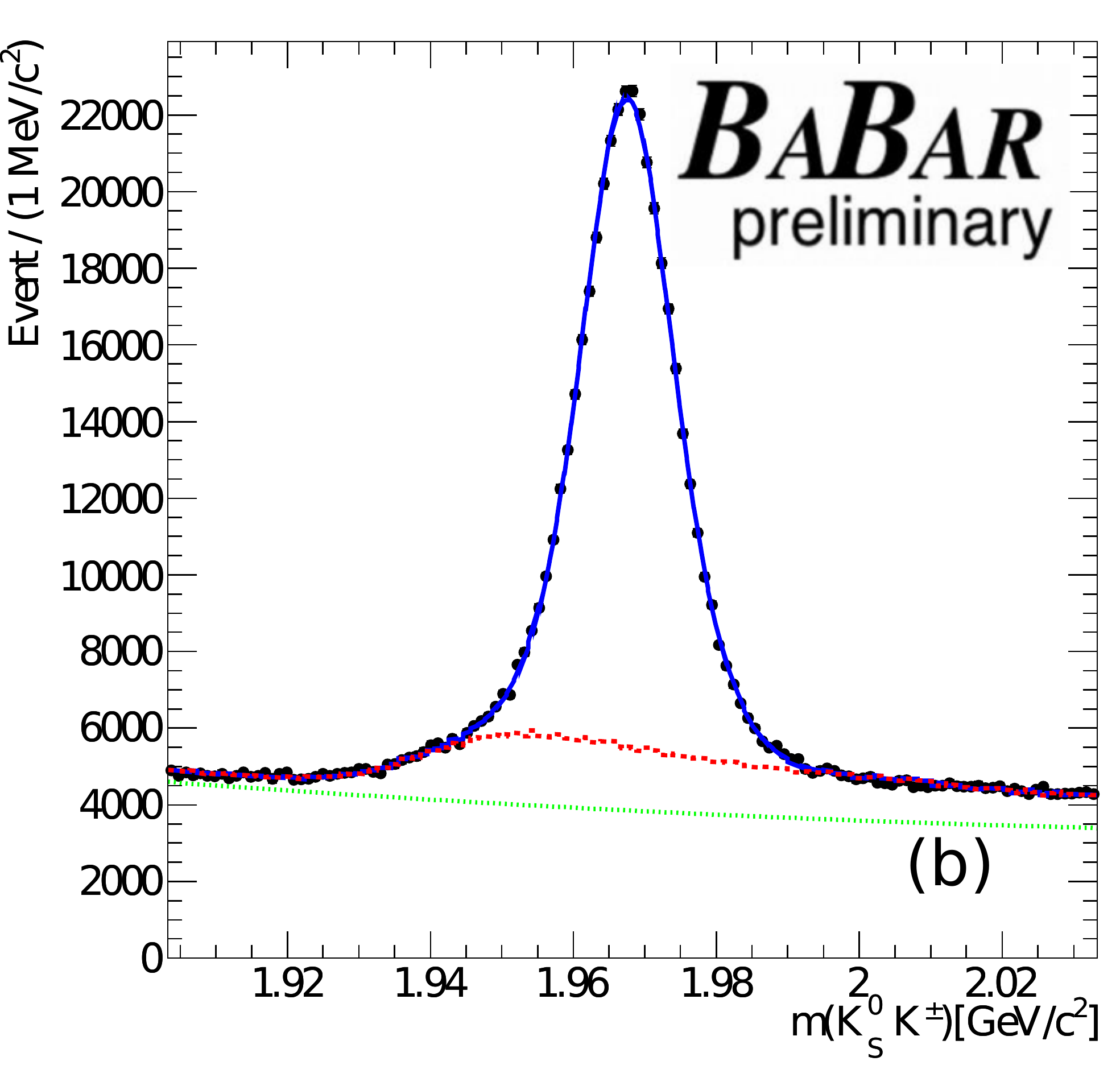} &
\includegraphics[width=0.31\textwidth]{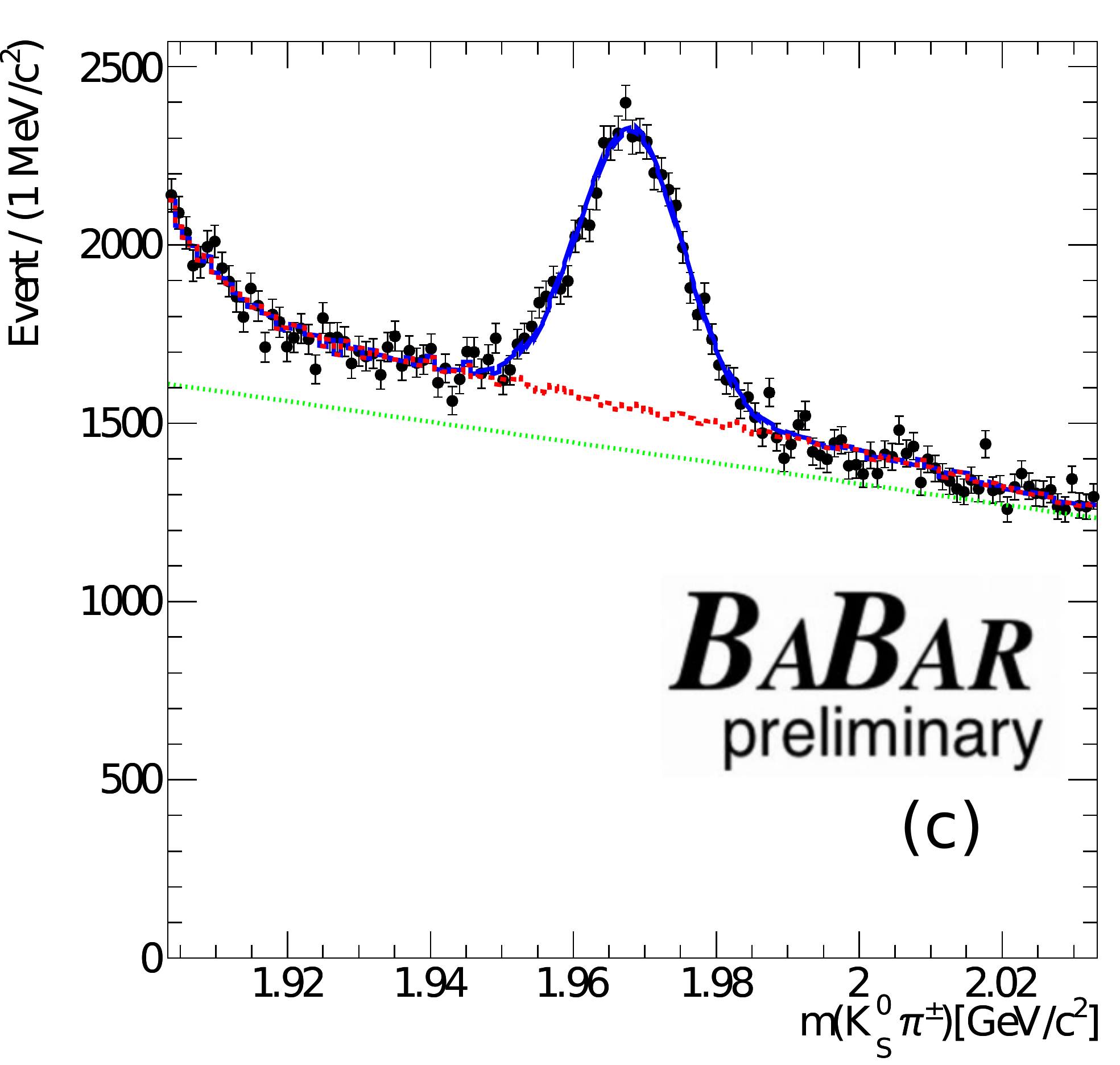} \\
\end{tabular}
\vspace{-0.3cm}
\caption{
  Invariant mass distribution for (a) \Dtoksk, (b) \Dstoksk, and (c) \Dstokspi
  candidates from the data (black points).
  In each figure, the solid blue curve shows the fit to the data,
  the dashed red curve is the sum of all backgrounds,
  while the dotted green curve represents combinatorial background only.
}
\label{fig1}
\vspace{-0.7cm}
\end{center}
\end{figure}

For each mode the signal yield is extracted using 
a binned maximum likelihood (ML) fit to the distribution 
of invariant mass, $m(\KS h)$, for the
selected $\Dps^\pm$ candidates.
The total probability density function (PDF) 
is the sum of signal and background components. 
The signal PDF is modeled as a sum of two Gaussian functions 
for both the \Dtoksk and \Dstoksk modes, and as a single Gaussian function for \Dstokspi mode.
The background PDF is taken as the sum of two components: background candidates from misidentified and 
partially reconstructed charm decays, and a combinatorial background component from other sources.
For \Dtoksk (\Dstokspi) the charm background is mainly from the tail of the invariant mass distribution of \Dstoksk (\Dtokspi) candidates.
For \Dstoksk the charm background originates from \Dtokspi where the \pipm is misidentified as \Kpm.
This wrong mass assignment shifts the reconstructed invariant mass to higher values,
and the resulting distribution is a broad peak with mean close to the $D_{s}^\pm$ mass.
The charm background distributions are modeled using a PDF sampled from the simulation events for these modes. 
The combinatorial background is described as a first(second)-order polynomial for \Dstokspi (\Dtoksk and \Dstoksk).
The fits to the $m(\KS h)$ distributions yield
$(159.4 \pm 0.8) \times 10^3$  \Dtoksk decays,
$(288.2 \pm 1.1) \times 10^3$  \Dstoksk decays, and
$(14.33 \pm 0.31) \times 10^3$ \Dstokspi decays.
The data and the fits are shown in Fig.~\ref{fig1}. 
All of the fit parameters are extracted from the fits to the data.

\section{$A_{\CP}$ Asymmetry Extraction}

For each channel, we determine $A_{\CP}$ by measuring the signal yield asymmetry $A$ defined as:
\begin{equation}
A=\frac{N_{\Dps^+}-N_{\Dps^-}}{N_{\Dps^+}+N_{\Dps^-}},
\end{equation}
where $N_{\Dps^+}$($N_{\Dps^-}$) is the number of $\Dps^+$($\Dps^-$) decays determined from the mass fit.
We consider $A$ to be the result of two other contributions in addition to $A_{\CP}$,
namely the forward-backward (FB) asymmetry ($A_{FB}$) and a detector-induced component.
We measure $A_{FB}$ together with $A_{\CP}$ using the selected dataset, 
while we correct the data for the detector-induced component
using coefficients from a control sample.

\begin{figure}[tb]
\begin{center}
\begin{tabular}{cc}
\includegraphics[width=0.47\textwidth]{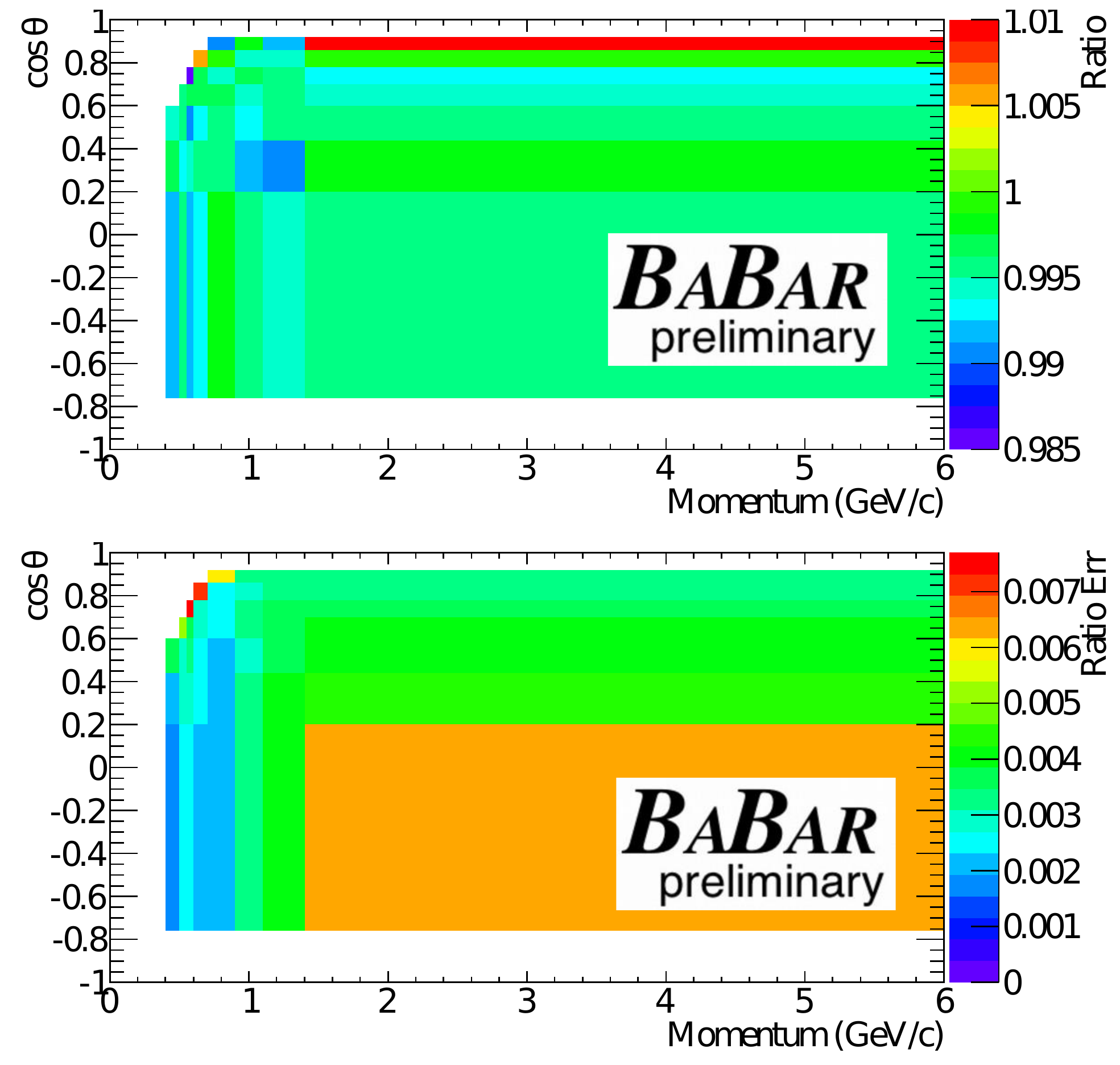} &
\includegraphics[width=0.47\textwidth]{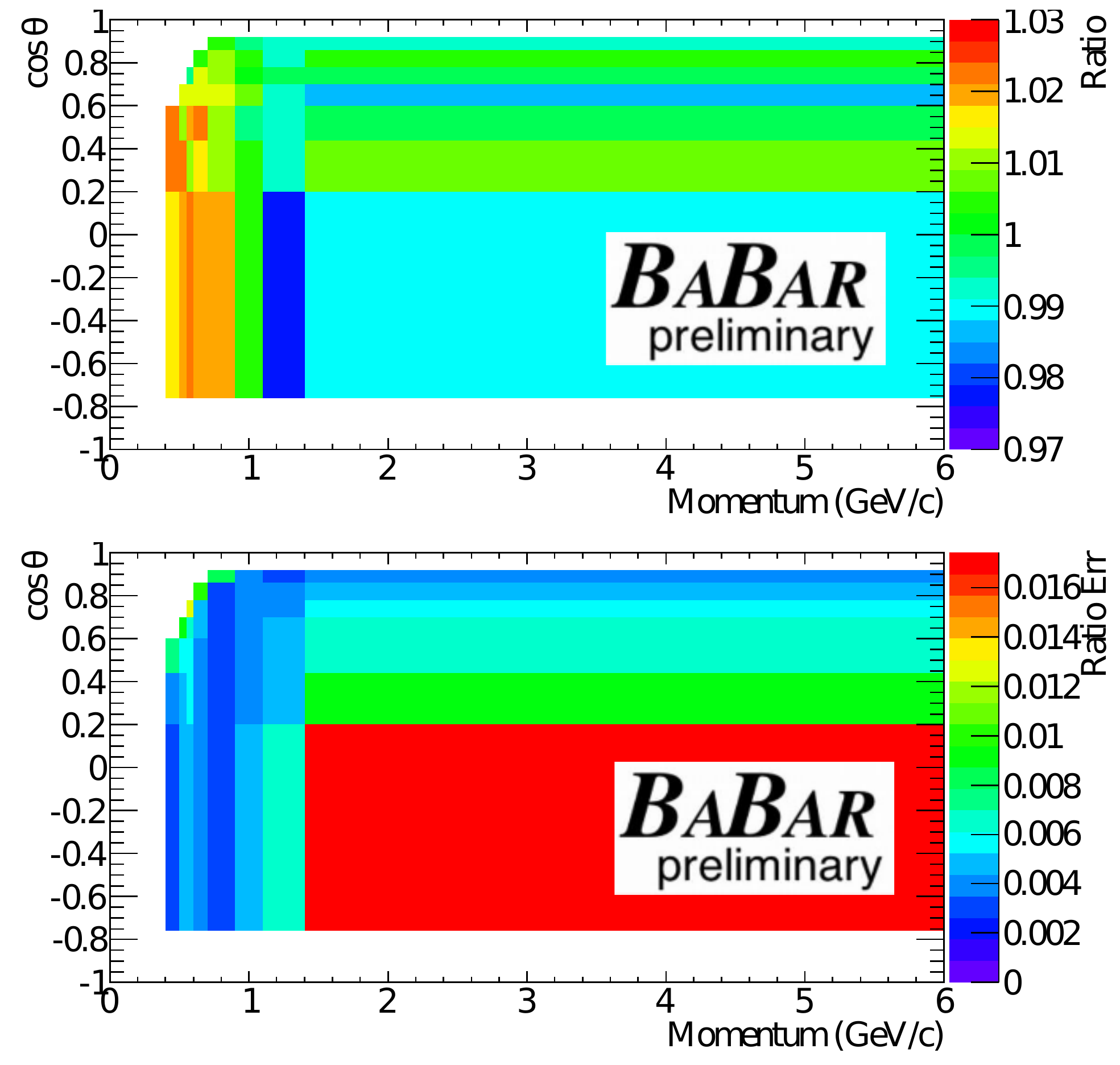} \\
\end{tabular}
\vspace{-0.3cm}
\caption{
  Ratio between the detection efficiency for positive and negative tracks (top) and
  corresponding statistical errors (bottom). 
  Plots on the left are for $\pi^+$ and $\pi^-$, while on the right for $K^-$ and $K^+$.
  The values are computed using the numbers of positive and negative tracks in the selected control sample.
}
\label{fig4}
\vspace{-0.7cm}
\end{center}
\end{figure}

In this analysis we use a data-driven method, described in detail in Ref.~\cite{delAmoSanchez:2011zza}, 
to determine the charge asymmetry in track reconstruction as a function of the magnitude of 
the track momentum and its polar angle. The method exploits the fact that the $\Upsilon (4S)\to \BB$ 
events provide sample of evenly populated positive and negative tracks, 
free of any physics-induced asymmetries, hence allowing the determination of detector-related asymmetries
in the reconstruction of charged-particle tracks.
Starting from a sample of $50.6\,\invfb$ of data collected at the $\Upsilon(4S)$ resonance and 
an off-resonance data set of $44.8\,\invfb$, we obtain a large sample of charged tracks and
apply the same charged pion or kaon track selection criteria 
used in the reconstruction of \Dpstoksk and \Dstokspi decays.
Then, by subtracting the off-resonance sample from the on-resonance sample,
we obtain a sample of more than 120 million pion candidates and 40 million kaon candidates, 
originating from $\Upsilon (4S)$ decays.
These are then used to compute the efficiency ratios for positive and negative kaons and pions.
The ratio values and associated statistical errors are shown in Fig.~\ref{fig4}.
For $\Dps^{-} \to \KS \Km$ ($D_s^{-} \to \KS \pim$) decays,
the $\Dps^{-}$ ($D_s^{-}$) yields, in intervals of kaon (pion) momentum and
$\cos\theta$, are weighted with the kaon (pion) efficiency ratios to
correct for the detection efficiency differences between \Kp and \Km (\pip
and \pim), leaving only FB and \CP asymmetries.
The largest correction is around 1\% for pions and 2\% for kaons.  

Neglecting the second-order terms that contain the product of $A_{\CP}$ and $A_{FB}$,
the resulting asymmetry can be expressed simply as the sum of the two.
Given that $A_{FB}$ is an odd function of $\cos\theta^*_D$, 
where $\theta^*_D$ is the polar angle of the $\Dps^\pm$ 
candidate momentum in the $\epem$ CM frame,
$A_{\CP}$ and $A_{FB}$ can be written as a function of $|\cos\theta^*_D|$ as follows:
\begin{align}
  A_{FB}(|\cos\theta^*_D|) &= \frac{A(+|\cos\theta^*_D|) - A(-|\cos\theta^*_D|)}{2} \\ 
  \intertext{and}
  A_{\CP}(|\cos\theta^*_D|) &= \frac{A(+|\cos\theta^*_D|) + A(-|\cos\theta^*_D|)}{2},
\label{eq:AcpAfb_intro}
\end{align}
where $A(+|\cos\theta^*_D|)$ is the measured asymmetry for the $\Dps^\pm$ candidates 
in a positive $\cos\theta^*_D$ bin and $A(-|\cos\theta^*_D|)$ in its negative counterpart.

A simultaneous Maximum Likelihood (ML) fit to the $\Dps^+$ and $\Dps^-$ invariant mass distributions
is carried out to extract the signal yield asymmetries for each one of the
ten equal intervals of $\cos\theta^*_D$, starting with
interval 0 defined as $-1.0 \ge \cos\theta^*_D \ge -0.8$.
The PDF shape that describes the distribution in each sub-sample
is the same as that used in the fit to the full sample,
but the following parameters are allowed to float separately
in each sub-sample (referred to below as split parameters): 
the yields for signal, charm background and combinatorial candidates; 
the asymmetries for signal and combinatorial candidates;
the width and the fraction of the Gaussian function with largest contribution to the signal PDF;
the first order coefficient for the polynomial of the combinatorial background.
For \Dtoksk the yield of charm background candidates has been fixed to 0 
in intervals 0, 1, and 2 in order to obtain a fully converging fit.
Interval 9 has the lowest number of candidates compared to the other intervals, so, 
for \Dtoksk and \Dstoksk, we use only one Gaussian function in this interval for the signal PDF
by fixing the fraction of the first Gaussian function to 1.
For the CPV asymmetry of charm background candidates 
we use the same floating parameters as for the signal candidates, 
because the largest source of CPV asymmetry for both samples is the CPV contribution from \KzKzb mixing. 
For the mode \Dstokspi, where the primary charm background channel, \Dtokspi, 
has the same magnitude but opposite sign asymmetry from \KzKzb mixing, we use a separate
parameter for the asymmetry of the charm background candidates. 
If the fit values for a split parameter are statistically compatible 
between two or more sub-samples, 
the parameter is forced to have the same floating value among those sub-samples only,
in order to achieve a more stable fit.
For \Dstokspi the width of the first Gaussian function for the signal PDF is set to the same
floating value for intervals 0, 1, 2, and 3.
The first order coefficient of the polynomial describing the combinatorial background
is set to the same floating value for intervals 3 to 7 (\Dtoksk), for intervals 4 to 7 (\Dstoksk),
and for intervals 1 to 6 (\Dstokspi).
The final fits for \Dtoksk, \Dstoksk, and \Dstokspi involve
a total of 70, 80 and 64 free parameters, respectively.

\begin{figure*}[tb] \begin{center}
\begin{tabular}{ccc}
\includegraphics[width=0.31\textwidth,clip=true]{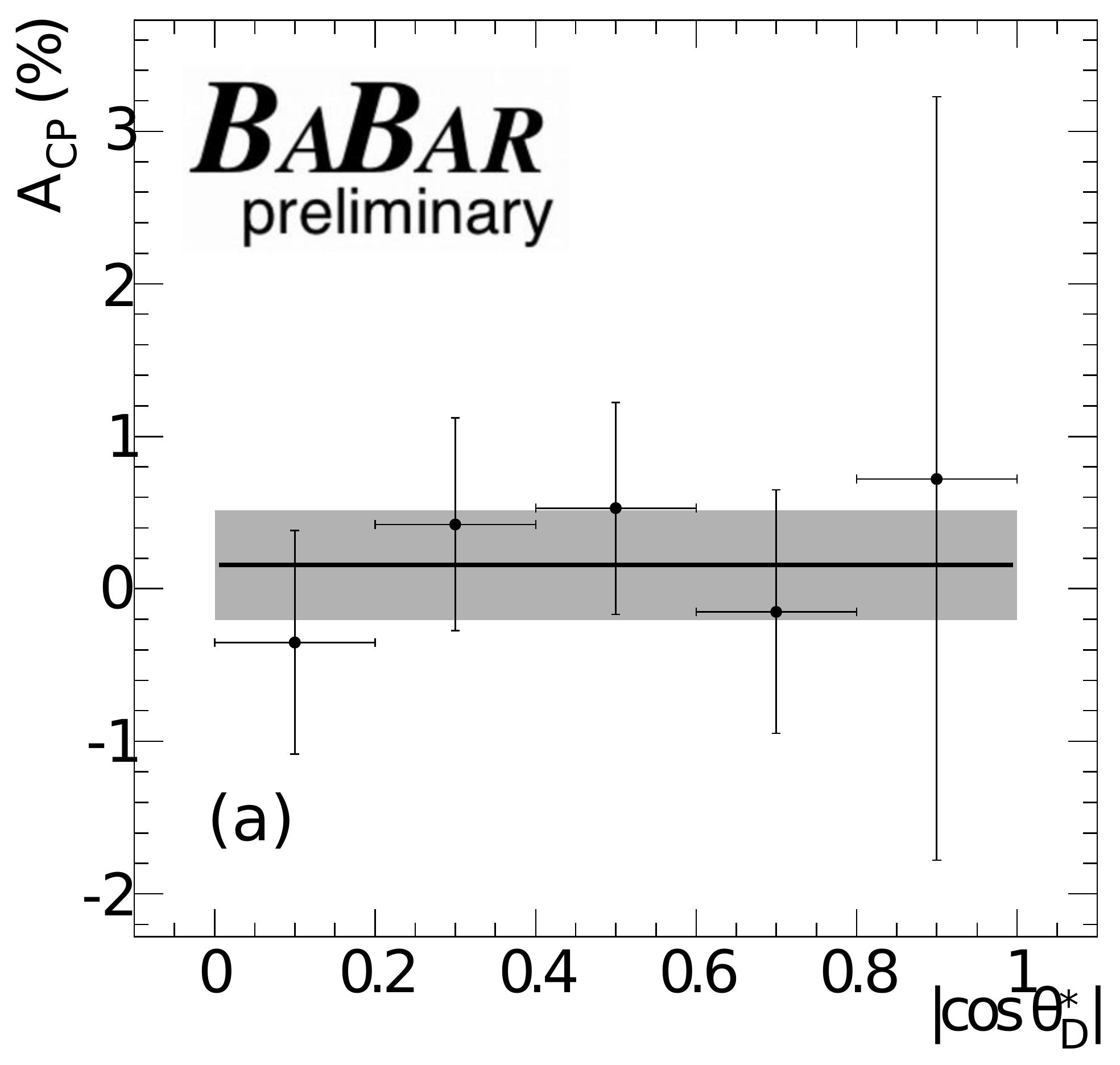} &
\includegraphics[width=0.31\textwidth,clip=true]{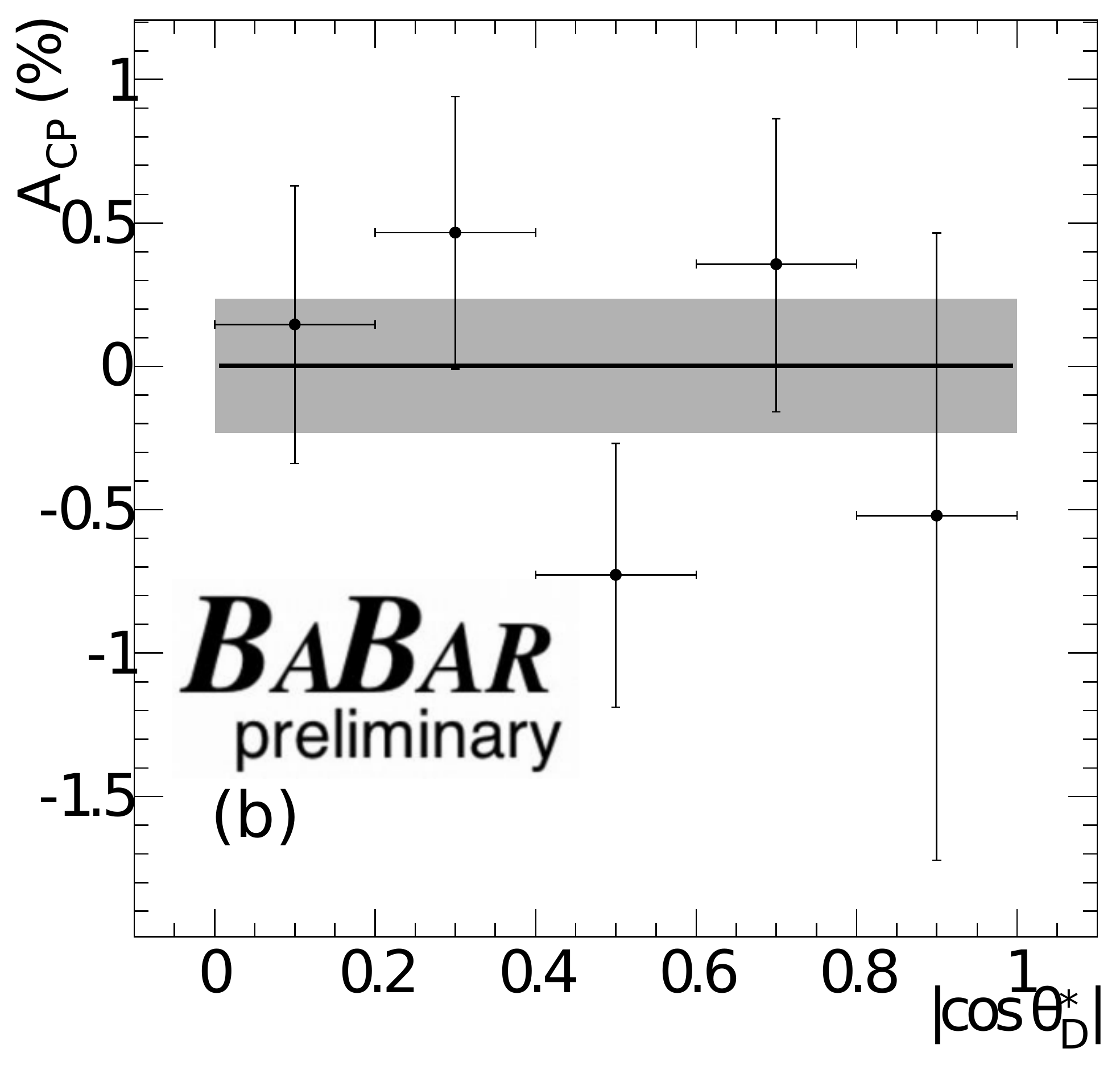} &
\includegraphics[width=0.31\textwidth,clip=true]{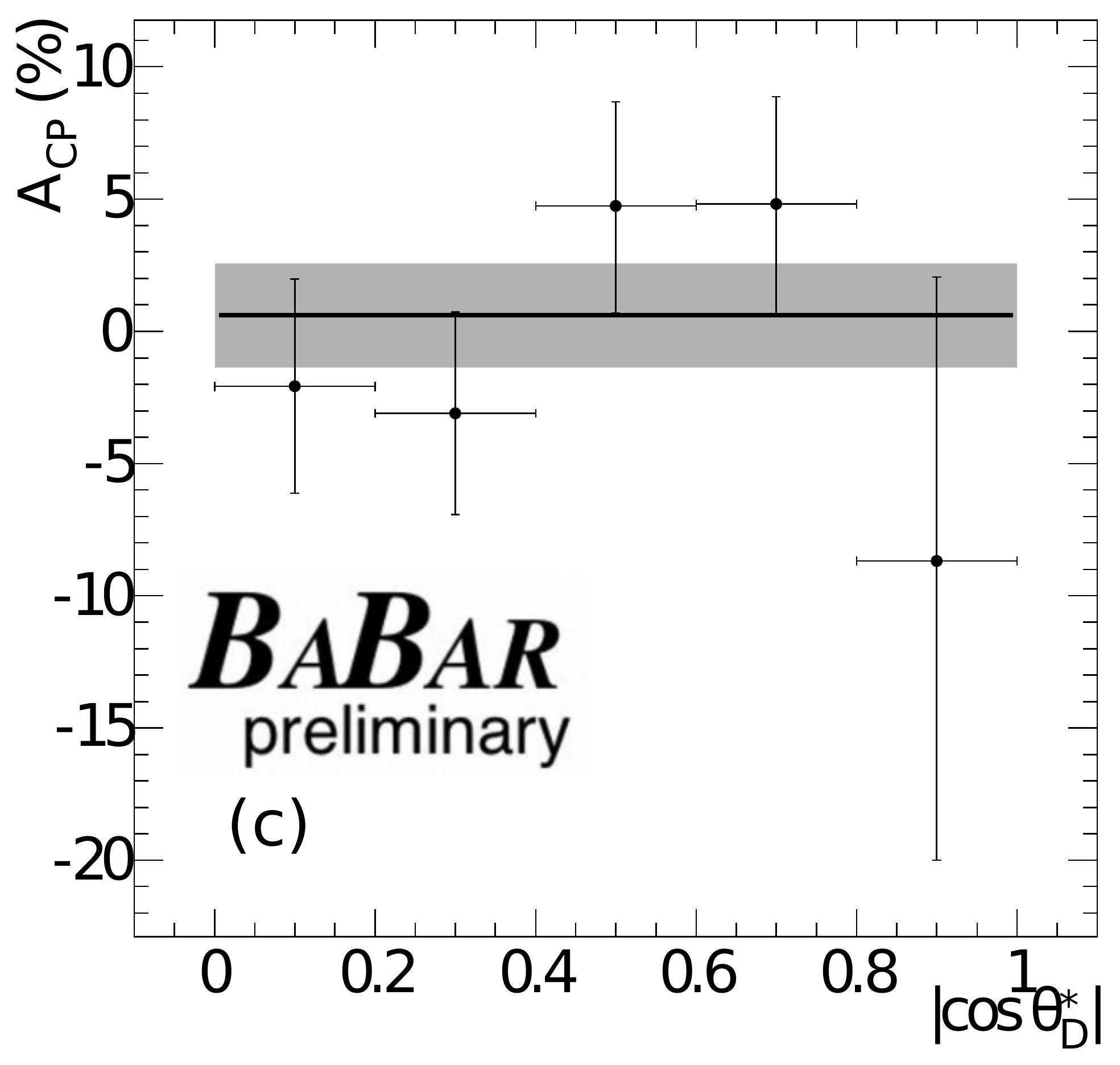} \\
\includegraphics[width=0.31\textwidth,clip=true]{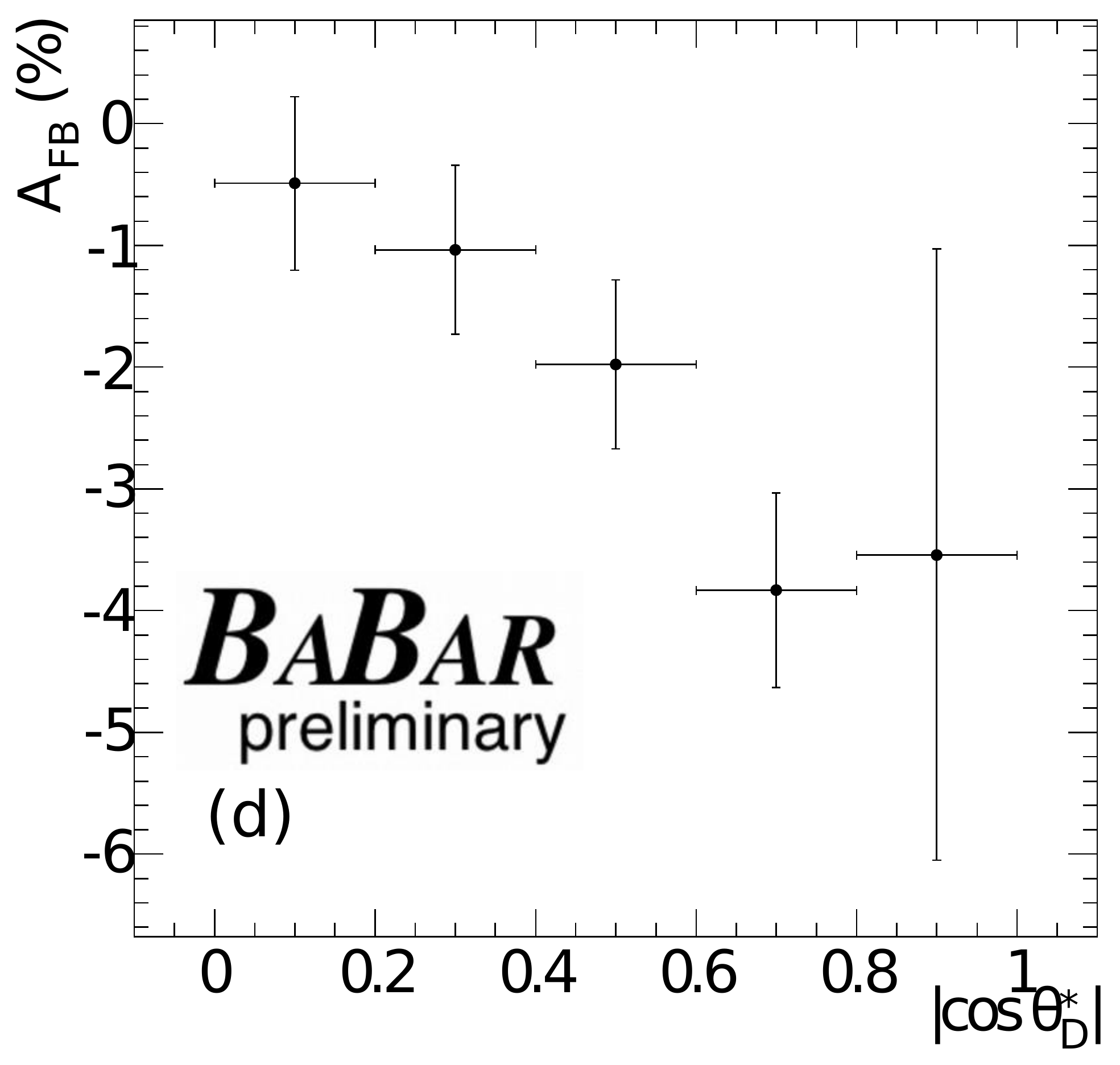} &
\includegraphics[width=0.31\textwidth,clip=true]{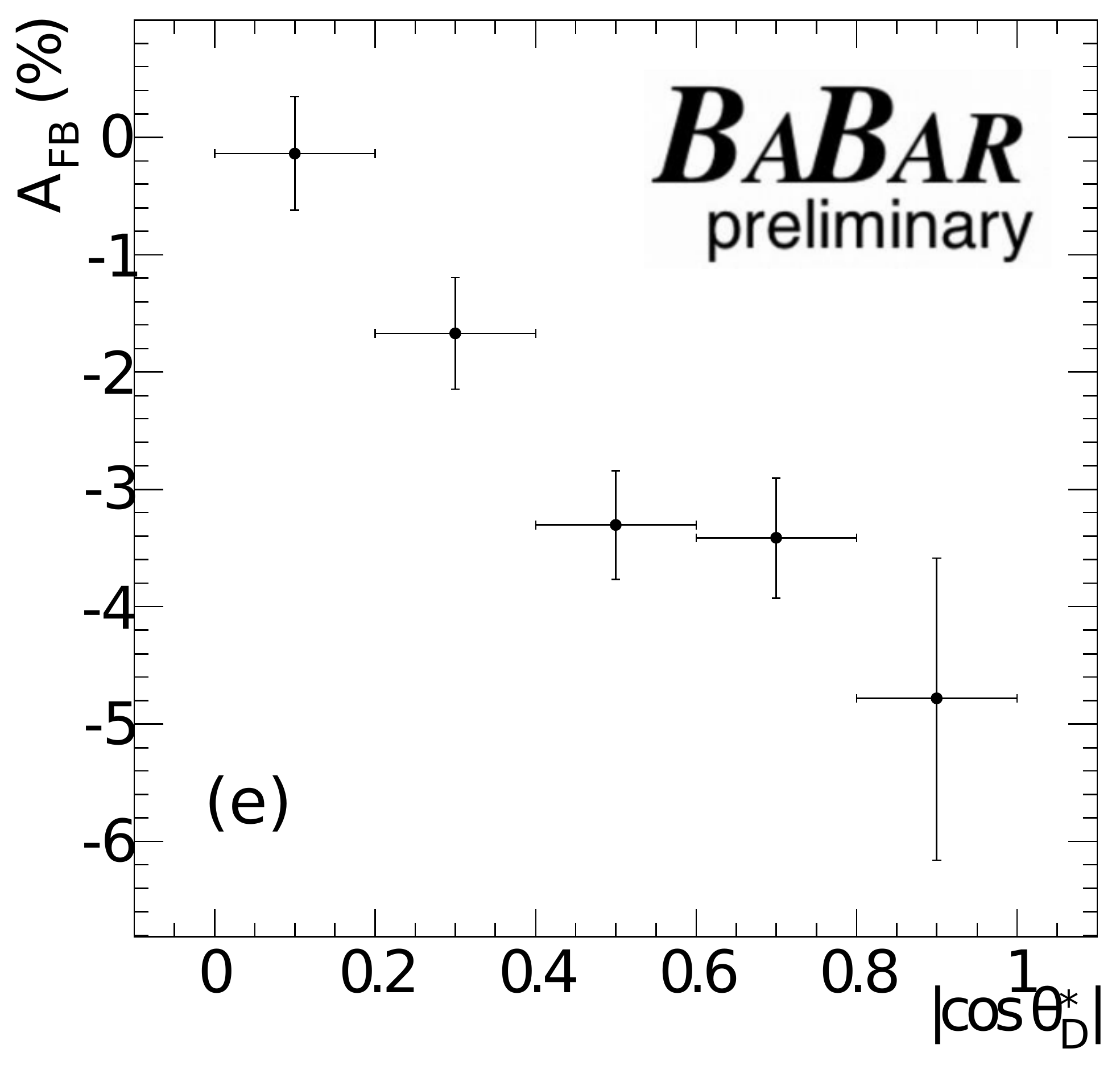} &
\includegraphics[width=0.31\textwidth,clip=true]{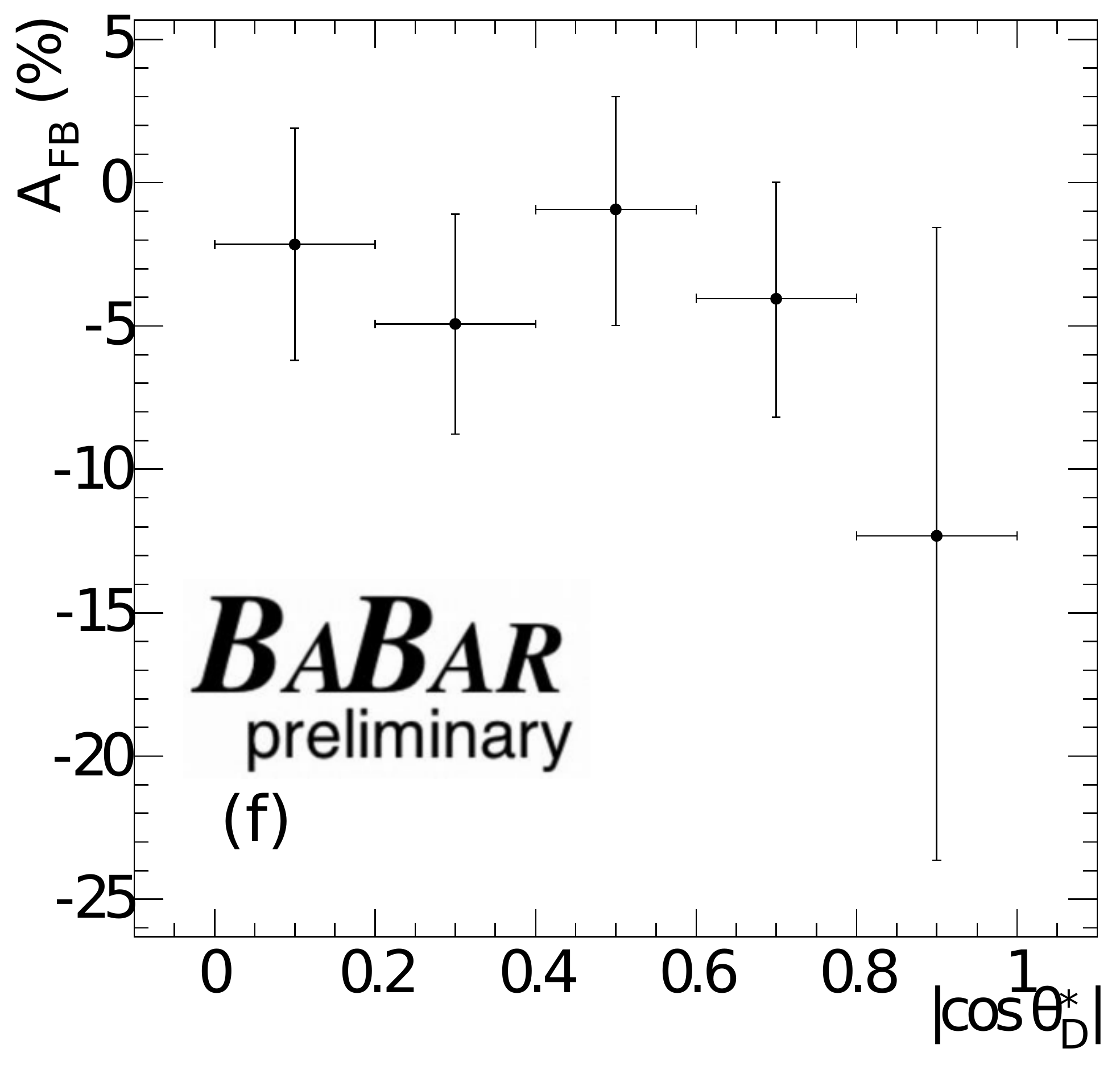} \\
\end{tabular}
\vspace{-0.3cm} \caption{
  $A_{\CP}$ asymmetry for (a) \Dtoksk, (b) \Dstoksk, and (c) \Dstokspi
  as a function of $|\cos\theta^*_D|$ in the data sample.
  The solid line represents the central value of $A_{\CP}$
  and the gray region is the $\pm1\,\sigma$ interval, both 
  obtained from a $\chi^2$ minimization
  assuming no dependence on $|\cos\theta^*_D|$.
  The corresponding $A_{FB}$ asymmetries are shown in (d), (e), and (f).
}
\label{fig6} \vspace{-0.7cm} \end{center} \end{figure*}

The $A_{\CP}$ and $A_{FB}$ values are shown in Fig.~\ref{fig6}.
The weighted average of the five $A_{\CP}$ values is found to be:
$(0.16 \pm 0.36)\%$ for \Dtoksk,
$(0.00 \pm 0.23)\%$ for \Dstoksk,
and $(0.6 \pm 2.0)\%$ for \Dstokspi,
where the errors are statistical only.

\section{Analysis Validation and Systematics}

We perform two tests to validate the analysis procedure for each channel. 
The first involves generating 5000 toy MC experiments using the PDF 
obtained from the fit to data, and then extracting $A_{\CP}$ from each experiment.
For the modes \Dtoksk and \Dstoksk, 
we find small biases of $-0.036 \pm 0.014$ and $+0.041 \pm 0.014$ 
(in units of statistical error), in the fitted values of the $A_{\CP}$ parameter.  
Therefore we apply a systematic correction of +0.013\% to the value of $A_{\CP}$
for \Dtoksk and of -0.01\% to that for \Dstoksk, to account for this effect.
The fit returns an accurate estimate of the statistical uncertainty for all the modes.
The second test involves fitting a large number of MC events from the 
full \babar detector simulation.
We measure $A_{\CP}$ from this MC sample to be within $\pm1\,\sigma$
of the generated value of zero.

A list of systematic error contributions and the quadrature sum for each mode
is reported in Table~\ref{tab_syst}. 
\begin{table*}[tb]
\caption{Summary of the systematic error contributions to the $A_{\CP}$ measurement in each mode.
The values are given as absolute percentage (\babar Preliminary).}
\begin{center}
\small
\begin{tabular}{|l|c|c|c|}\hline
Systematic uncertainty [\%] & \Dtoksk & \Dstoksk & \Dstokspi \\
\hline
Efficiency of PID selectors          & 0.05\% & 0.05\% & 0.05\% \\
Statistics of the control sample  & 0.23\% & 0.23\% & 0.06\% \\
Mis-identified tracks in the control sample    & 0.01\% & 0.01\% & 0.01\% \\
Binning in $\cos\Theta$              & 0.04\% & 0.02\% & 0.27\% \\
\KzKzb regeneration                & 0.05\% & 0.05\% & 0.06\% \\
\KSKL interference                & 0.015\% & 0.014\% & 0.008\% \\
\hline
Total                                & 0.25\% & 0.24\% & 0.29\% \\
\hline
\end{tabular}
\normalsize
\label{tab_syst}
\end{center}
\end{table*}
Primary sources of systematic uncertainty are the statistical uncertainties 
in the detection efficiency ratios used to weight the $\Dps^\pm$ yields and 
the contributions from misidentified particles in the data control sample used 
to determine the charge asymmetry in track reconstruction efficiency.

The technique used here to remove the charge asymmetry 
from detector-induced effects produces a small systematic uncertainty in 
the measurement of $A_{CP}$ due to the statistical error on the relative efficiency estimation.
We perform simultaneous fits to 500 samples of selected \Dstoksk candidates
where the applied corrections are smeared according to their errors, 
to produce a distribution of the deviations of $A_{\CP}$ values from the nominal one. 
The standard deviation of this distribution is 0.23\%, and it is assigned as systematic
uncertainty to the \Dstoksk and \Dtoksk modes, 
because this contribution depends only on the type of charged particle in the mode itself.
For the \Dstokspi mode we assign a systematic error contribution obtained as in Ref.~\cite{delAmoSanchez:2011zza}.
This is the dominant source of systematic uncertainty for the \Dpstoksk modes, 
as shown in Table~\ref{tab_syst}.

The small fraction of mis-identified particles in the generic track sample 
can introduce small biases in the estimation of the efficiencies, 
and subsequently in the $A_{\CP}$ measurements.
Because of the good agreement between data and MC samples, we can use the MC-simulated candidates 
to measure the shift in the $A_{\CP}$ value from the fit when the corrections are, or are not, applied.
Again, this contribution depends only on the type of charged-particle track, hence 
for the \Dstokspi mode we assume the same shift as that obtained 
in Ref.~\cite{delAmoSanchez:2011zza}, namely +0.05\%. 
Fitting the \Dstoksk MC sample when the corrections are, or are not, applied,
we obtain a shift of +0.05\%, and we assume this same uncertainty for the \Dstoksk and \Dtoksk modes.
As a result, for each mode we shift the measured $A_{\CP}$ value 
by the net systematic shift to correct for this bias,
and then, conservatively, include the same value as a contribution to the systematic uncertainty.

Using MC simulation, we evaluate an additional systematic uncertainty
of $\pm0.01\%$ due to a possible charge asymmetry present in the control sample 
before applying the selection criteria.
Another systematic uncertainty from the simultaneous ML fit is due 
to the choice of interval-size in $\cos\theta^*_D$. This can be estimated
using the largest $A_{\CP}$ deviation when the fit is performed using 8 or 12 intervals
of $|\cos\theta^*_D|$, instead of 10.
This is the dominant source of systematic uncertainty for the \Dstokspi mode, 
as shown in Table~\ref{tab_syst}.

We also consider a possible systematic uncertainty due to the regeneration
of neutral kaons in the material of the detector,
since \Kz and \Kzb mesons produced in the decay process can interact with the 
material in the tracking volume before they decay.
Following a method similar to that described in Ref.~\cite{Ko:2010mk},
we compute the probability for a \Kz or \Kzb to interact 
inside the \babar tracking system,
and estimate an associated systematic uncertainty of $0.05-0.06\%$.

Another systematic contribution can be generated by the interference
between the amplitudes of intermediate \KS and \KL, because we are reconstructing
our \KS using $\KS\to\pi^+\pi^-$, which is not a pure \KS amplitude~\cite{Grossman:2011zk}.
The bias in the asymmetry contribution induced by \KzKzb mixing
depends directly on the \KS reconstruction efficiency as a function of the \KS proper time
and, approximately, the more the efficiency is not constant, the larger becomes the bias.
We produce the \KS reconstruction efficiency distribution
as a function of the proper time using MC truth-matched events after the full selection.
Then, following the method in Ref.~\cite{Grossman:2011zk}, 
we estimate the asymmetry correction factor $\Delta A_{\CP}$ defined as:
\begin{equation}
\Delta A_{\CP} = A_{\CP}^{\textrm{corr}} - A_{\CP}^{\textrm{fit}},
\end{equation}
where $A_{\CP}^{\textrm{fit}}$ is the value obtained from the fit and
$A_{\CP}^{\textrm{corr}}$ is the corrected value.
The correction factors are reported in Table~\ref{tab_final}
and, to be conservative, we include their absolute values
as a contribution to the systematic uncertainty.
In a similar mode, we also estimate the correction factor for \Dtokspi mode 
using the \KS reconstruction efficiency distribution after the selection detailed 
in Ref.~\cite{delAmoSanchez:2011zza} and we find a value of +0.002\%.
All these corrections are rather small, even compared to those estimated in a
similar analysis~\cite{BABAR:2011aa}. 
The smaller values of the corrections in this analysis are due to the 
improved efficiency for \KS mesons with short decay times that we obtain 
by making a requirement on the decay length divided by its uncertainty 
rather than on the decay length alone.

\section{Conclusions and Acknowledgements}

In conclusion, we measure the direct \CP asymmetry, $A_{\CP}$, in
\Dtoksk, \Dstoksk, and \Dstokspi decays using approximately 
159,000, 288,000, and 14,000 signal candidates, respectively.
The measured $A_{\CP}$ value for each mode is reported in Table~\ref{tab_final},
where the first error is statistical and the second is systematic.  
In the last row of the table, we also report the $A_{\CP}$ values after 
subtracting the expected $A_{\CP}$ contribution in each mode due to \KzKzb mixing.
\begin{table*}[tb]
\caption{Summary table for $A_{\CP}$ measurements.
Uncertainties, where reported, are first statistical, and second systematic
(\babar Preliminary).} 
\begin{center}
\footnotesize
\begin{tabular}{|l|c|c|c|}
\hline
  & \Dtoksk & \Dstoksk & \Dstokspi \\
\hline \hline
$A_{\CP}$ value from the fit      & $(0.16 \pm 0.36)\%$ & $(0.00 \pm 0.23)\%$ & $(0.6 \pm 2.0)\%$ \\ 
\hline \hline
Bias Corrections & \multicolumn{3}{|c|}{}\\
\hline 
Toy MC experiments & $+0.013\%$ & $-0.01\%$ & $-$ \\ 
PID selectors & $-0.05\%$ & $-0.05\%$ & $-0.05\%$ \\
\KSKL interference 
& $+0.015\%$ & $+0.014\%$ & $-0.008\%$ \\
\hline \hline
$A_{\CP}$ corrected value & $(0.13 \pm 0.36 \pm 0.25)\%$ & $(-0.05 \pm 0.23 \pm 0.24)\%$ & $(0.6 \pm 2.0 \pm 0.3)\%$\\
\hline \hline
$A_{\CP}$ contribution & \multirow{2}{*}{$(-0.332 \pm 0.006)\%$} & \multirow{2}{*}{$(-0.332 \pm 0.006)\%$} & 
\multirow{2}{*}{$(0.332 \pm 0.006)\%$} \\
from \KzKzb mixing &&&\\
\hline \hline
$A_{\CP}$ value (charm only) & $(0.46 \pm 0.36 \pm 0.25)\%$ & $(0.28 \pm 0.23 \pm 0.24)\%$ & $(0.3 \pm 2.0 \pm 0.3)\%$\\
\hline
\end{tabular}
\normalsize
\label{tab_final}
\end{center}
\end{table*}
These results are consistent with zero and with the SM prediction 
within one standard deviation.

We are grateful for the 
extraordinary contributions of our \pep2 colleagues in
achieving the excellent luminosity and machine conditions
that have made this work possible.
The success of this project also relies critically on the 
expertise and dedication of the computing organizations that 
support \babar.
The collaborating institutions wish to thank 
SLAC for its support and the kind hospitality extended to them. 
This work is supported by the
US Department of Energy
and National Science Foundation, the
Natural Sciences and Engineering Research Council (Canada),
the Commissariat \`a l'Energie Atomique and
Institut National de Physique Nucl\'eaire et de Physique des Particules
(France), the
Bundesministerium f\"ur Bildung und Forschung and
Deutsche Forschungsgemeinschaft
(Germany), the
Istituto Nazionale di Fisica Nucleare (Italy),
the Foundation for Fundamental Research on Matter (The Netherlands),
the Research Council of Norway, the
Ministry of Education and Science of the Russian Federation, 
Ministerio de Ciencia e Innovaci\'on (Spain), and the
Science and Technology Facilities Council (United Kingdom).
Individuals have received support from 
the Marie-Curie IEF program (European Union) and the A. P. Sloan Foundation (USA).


\begin{thebibliography}{99}
\bibitem{CPTinv}
The contribution from \CP violation in $B$ decays from the Standard Model
processes is estimated to be negligible.
\end{thebibliography}
\end{document}